\documentclass[reprint,aps,prd,twocolumn,notitlepage,nofootinbib,showpacs,preprintnumbers,superscriptaddress]{revtex4-1}
\usepackage[T1]{fontenc}
\usepackage[utf8]{inputenc}
\usepackage{bm}
\usepackage{amsmath}
\usepackage{amssymb}
\usepackage{esint}
\usepackage{color}
\usepackage{graphicx}
\PassOptionsToPackage{normalem}{ulem}
\usepackage{ulem}

\usepackage{times}
\usepackage{hyperref}

\newcommand{\be}{\begin{equation}}
\newcommand{\ee}{\end{equation}}
\newcommand{\ba}{\begin{eqnarray}}
\newcommand{\ea}{\end{eqnarray}}

\def\pa{\partial}

\def\a{\alpha}
\def\b{\beta}
\def\g{\gamma}

\def\d{\delta}

\def\e{\epsilon}

\def\m{\mu}
\def\n{\nu}

\def\r{\rho}

\def\s{\sigma}

\def\o{\omega}

\begin{document}
\preprint{CCQCN-2015-109}
\title{Unifying Ghost-Free Lorentz-Invariant Lagrangians}

\author{Wenliang LI}
\email{lii.wenliang@gmail.com}
\affiliation{
APC, Universit\'e Paris 7, CNRS/IN2P3, CEA/IRFU, Obs. de Paris, Sorbonne Paris Cit\'e, B\^atiment Condorcet, 
10 rue Alice Domon et L\'eonie Duquet, F-75205 Paris Cedex 13, France 
({UMR 7164 du CNRS})}
\affiliation{
Crete Center for Theoretical Physics (CCTP) and Crete Center for Quantum Complexity and
Nanotechnology (CCQCN), Department of Physics, University of Crete, P.O. Box 2208, 71003, Heraklion, Greece}

\date{October 19, 2015}

\begin{abstract}
We present the details of the novel framework for Lagrangian field theories 
that are Lorentz-invariant
and lead to at most second order equations of motion. 
The use of antisymmetric structure is of crucial importance. 
The general ghost-free Lagrangians are constructed and then translated into the language of differential forms. 
The ghost-freeness has a geometric nature. 
A novel duality is proposed which generalizes the Hodge duality in Maxwell's theory. 
We discuss how the well-established theories are reformulated and propose many new theories. 
\end{abstract}

\pacs{11.10.-z, 11.30.-j, 04.50.-h, 02.40.-k }

\maketitle

\section{Introduction}\label{sec-into}
Antisymmetry plays a fundamental role in theoretical physics. 
In Lagrangian field theories, the field strengths in gauge theories and 
the Riemann curvature tensor in general relativity are the commutators of two covariant derivatives. 
By Legendre transformation, the Lagrangian theories are reformulated as Hamiltonian systems. 
Both the symplectic structures of the phase space and the Poisson brackets in Hamilton's equations are antisymmetric objects. 
In addition, the Poisson brackets are closely connected to the commutator brackets in quantum mechanics. 

Besides general relativity, antisymmetric structure appears in the theories of modified gravity. 
Since Albert Einstein conceived the general theory of relativity, 
there have been many attempts to construct gravitational theories that are different from Einstein's theory. 

General relativity can be thought of as a nonlinear theory of massless spin-2 field. 
One of the simplest idea of modifying the theory of gravity is to give the graviton a mass. 
To avoid the ghost-like degrees of freedom, the mass term in the linear theory needs the Fierz-Pauli tuning \cite{Fierz:1939ix} 
\be
\mathcal L_{\text{FP}}\sim m^2\,({h_{\m_1}}^{\m_1}{h_{\m_2}}^{\m_2}-{h_{\m_1}}^{\m_2}{h_{\m_2}}^{\m_1})
=m^2\,{h_{\m_1}}^{[{\m_1}}{h_{\m_2}}^{{\m_2}]},
\ee
where $[\dots]$ indicates the indices inside the bracket are antisymmetrized. 
We use the unnormalized convention for antisymmetrization throughout this paper. 

Recently, the linear ghost-free mass term was extended to the nonlinear level \cite{deRham:2010kj} 
\be
\mathcal L^{(k)}_{\text{dRGT}}\sim m^2\, {(\sqrt{g\,\eta^{-1}}\,)_{\m_1}}^{[\m_1}\dots {(\sqrt{g\,\eta^{-1}}\,)_{\m_k}}^{\m_k]},\label{dRGT}
\ee
where $\sqrt{g\,\eta^{-1}}$ is the square root of the metric product $g\,\eta^{-1}$. 
At the lowest order, the dRGT terms reduce to the Fierz-Pauli mass term.

Another possibility is to consider high order derivative theories of the metric field. 
Lovelock showed in \cite{Lovelock:1971yv}  that the general Lagrangian of metric theories that lead to second order equations of motion is a sum of the terms below
\be
\mathcal L_{\text{Lovelock}}^{(k)}
\sim\sqrt{-g}\,{R_{\m_1\m_2}}^{[\m_1\m_2}\dots{R_{\m_{2k-1}\m_{2k}}}^{\m_{2k-1}\m_{2k}]}.\label{Lovelock}
\ee
The Lovelock terms are the products of the Riemann curvature tensor whose indices are contracted antisymmetrically. 

The third choice to modify gravity theory is introducing more degrees of freedom, 
for example an additional scalar field in Brans-Dicke theory \cite{Brans:1961sx}. 
Following Lovelock's approach, the most general scalar-tensor theory whose equations of motion are directly of second order was constructed by Horndeski \cite{Horndeski:1974wa}. 
Let us consider flat spacetime for simplicity. 
The ghost-free Lagrangians of single scalar field with high order derivative terms are known as Galileons \cite{Nicolis:2008in} 
\be
\mathcal L^{(k)}_{\text{Galileon}}\sim f_k\,\pa_{\m_1}\pa^{[\m_1}\Phi\dots \pa_{\m_k}\pa^{\m_k]}\Phi,\label{Gali}
\ee
where $f_k$ are scalar functions of $\Phi$ and $\pa_\r\Phi\,\pa^\r\Phi$. 

There are some common features in these Lagrangian theories of modified gravity (\ref{dRGT},\,\ref{Lovelock},\,\ref{Gali}) :
\begin{enumerate}
\item
they are Lorentz-invariant;
\item
they are free of ghost-like degrees of freedom;
\item
the indices are contracted antisymmetrically;
\item
the numbers of possible terms are finite.
\end{enumerate}
The precise meanings of Lorentz-invariant and ghost-free are discussed in section \ref{sec-genL}. 
The fourth feature is a consequence of the third one 
because the numbers of antisymmetrized indices can not be larger than the dimensions of spacetime. 
Are the first two features related to the third one as well? 

In addition, the first three features are not limited to the theories of modified gravity. 
They are shared by the conventional theories. 
For example, the linearized Einstein-Hilbert term reads
\be
\mathcal L_{\text{EH}}\sim {h_{\m_1}}^{[{\m_1}}\pa_{\m_2}\pa^{\m_2} {h_{\m_3}}^{{\m_3}]}
\ee
and the Maxwell kinetic term for spin-1 field is
\be
\mathcal L_{\text{Maxwell}}\sim A_{\m_1} \pa_{\m_2}\pa^{[{\m_1}}A^{{\m_2}]}.
\ee

Based on these examples, we are tempted to believe that there exists a unifying framework for Lagrangian field theories that are both ghost-free and Lorentz-invariant. 
Antisymmetrization should be the key ingredient. 
The goal of this work is to present such a general framework \cite{Li:2015vwa} in detail. 

This paper is organized as follows: 
in section \ref{sec-genL} the general ghost-free Lagrangians are ``derived" and 
the importance of antisymmetric Kronecker delta is discussed; 
in section \ref{sec-form} these Lagrangians are translated into the language of differential forms, the geometric nature of ghost-freeness is elucidated 
and a generalization of the Hodge duality is proposed; 
in section \ref{sec-ex}, examples are presented which include the reformulations of the known theories 
and the proposals of new ghost-free theories; 
in section \ref{sec-ccl} we summarize the results and discuss the outlook. 

The appendices are some concrete examples of the abstract discussions in the main text. 
They are a simple example of Ostrogradsky's instability in Appendix \ref{app-Ost}, 
linearized (massive) gravity as ghost-free theories of spin-2 fields in Appendix \ref{app-lins2}, 
the translation of the Galileons to differential forms in Appendix \ref{app-form}, 
the Maxwell term and the Einstein-Hilbert terms as double forms in Appendix \ref{app-douf}. 

\section{General ghost-free Lagrangians}\label{sec-genL}
In this section, the general framework is developed in the language of tensors. 
The general expression of ghost-free Lagrangians is presented in subsection \ref{subs-genL}. 
The basic assumptions are Lorentz-invariance and second order equations for the decomposed fields, 
which are explained in subsections \ref{subs-Li}, \ref{subs-Ost} and \ref{subs-lon}. 
The key ingredient is the antisymmetric Kronecker delta, 
which is based on two chains of antisymmetrized indices constructed in subsections \ref{subs-chains} and \ref{subs-tenind}. 
Some technical points are discussed in subsections \ref{subs-1order} and \ref{subs-sindex}. 

\subsection{Lorentz invariance}\label{subs-Li}
In a Lorentz-invariant Lagrangian, the dynamical fields transform covariantly as Lorentz tensors under global, external Lorentz transformation. 
The indices are raised and lowered by the Minkowski metric. 

In the metric theories, the dynamical field can be identified as the full metric or the metric perturbation around the Minkowski background
\be
h_{\m\n}=g_{\m\n}-\eta_{\m\n},
\ee
where $g_{\m\n}$ and $\eta_{\m\n}$ are the full and Minkowski metrics. 
A theory of spin-2 field is Lorentz-invariant if it is constructed from 
$h_{\m\n},\,\eta_{\m\n},\,\eta^{\m\n}$ and $\pa_\m$ with no free index.

In general, the indices in a Lorentz-invariant Lagrangian are contracted with a constant Lorentz-covariant tensor. 
The building blocks of this constant tensor are the Minkowski metric $(\eta_{\m\n},\,\eta^{\m\n})$ 
and the Levi-Civita symbol
\ba
\epsilon_{\m_1\dots\m_D}=&\,\delta_{[\m_1}^0\dots \delta_{\m_D]}^{D-1}=(-1)^p,
\ea
where $p$ is the number of exchanges of indices required to transform $\{\m_1,\dots,\m_D\}$ to $\{0,\,1,\dots,D-1\}$. 
$D$ is the spacetime dimension. 
By definition, the Levi-Civita symbol is totally antisymmetric in all the indices
\be
\epsilon_{\dots\m_i\dots\m_j\dots}=-\epsilon_{\dots\m_j\dots\m_i\dots}.
\ee

In this work, we consider Lorentz-invariant theories 
which satisfy the above requirements. 

\subsection{Ostrogradsky's instability}\label{subs-Ost}
The Hamiltonian of a theory should be bounded from below. 
If a kinetic term has a wrong sign, the energy will not have a lower bound 
and the modes with negative kinetic energy are ghost-like degrees of freedom. 

When a Lagrangian contains high order terms, 
the Hamiltonian could be linear in one momentum 
and therefore not bounded from below. 
This is known as the Ostrogradsky instability \cite{Woodard:2006nt} . 
Here ``high order terms'' indicates second and higher order derivative terms. 
To illustrate the instability induced by higher order terms in the Lagrangian, 
an example is discussed in Appendix \ref{app-Ost}. 
In this example, a fourth order term leads to additional degrees of freedom 
whose kinetic term has a wrong sign. 

However, if the higher order Lagrangian is degenerate, the equations of motion could be of second order. 
In this case, if the combination of higher order terms has a correct overall sign, 
the Hamiltonian could be bounded from below and the ghost-like degrees of freedom are eliminated. 
The absence of higher order terms in the equations of motion is one of the main working assumptions in this work
\footnote{Recently, there appeared new classes of scalar-tensor theories \cite{Gleyzes:2014dya} and \cite{BHorndeski-XG}, 
which lead to higher order equations of motion but still seem be free of ghost-like degrees of freedom. 
Hamiltonian analyses in the unitary gauge \cite{BHHam} support the healthiness of these new theories. 
In \cite{Deffayet:2015qwa}, it was shown that, 
in (generalized) Galileon models with arbitrary coefficients, 
the third order time derivative terms arising in the equations of motion can be removed 
by using the time derivative of second order equations of motions. 
In this paper, we will not consider this interesting possibility and require the equations of motion arising from the Lagrangians are directly of second order.}
.

\subsection{Longitudinal modes}\label{subs-lon}
In the theories of tensor fields, ghost-like degrees of freedom could be propagating 
even if the equations of motion are of second order. 

For example, the unique ghost-free Lorentz-invariant kinetic terms for spin-1 field is the Maxwell kinetic term, 
and the unique ghost-free mass term for spin-2 field requires the Fierz-Pauli tuning. 
If one relaxes the relative coefficients of the Lorentz-invariant terms, 
the kinetic terms for the spin-1 field and the mass term for the spin-2 field 
will generate ghost-like degrees of freedom. 

The reason behind this is that the equations of motion for the longitudinal modes could be of higher order. 
For example, we can decompose the spin-2 field into the transverse and longitudinal modes
\be
h_{\m\n}=h_{\m\n}^T+\pa_\m A_\n+\pa_\n A_\m+\pa_\m\pa_\n\phi,\label{spin2-de}
\ee
\be
\pa^\m h_{\m\n}^T=\pa^\m A_\m=0,\label{trans}
\ee
and a generic two-derivative action could lead to fourth order terms of $A_\m$ and sixth order terms of $\Phi$ 
in the equations of motion. 

Since some of the tensor indices become derivative indices, 
the longitudinal modes are more dangerous than the transverse modes. 
Among them, the longitudinal scalar modes are the most dangerous because all the tensor indices are transformed into derivative indices. 

For tensor fields, the ghost-free working assumption becomes that 
the equations of motion for the decomposed modes are at most of second order. 

In Appendix \ref{app-lins2}, the linear theories of spin-2 fields are investigated. 
The structures of the kinetic term and the mass term are uniquely determined 
by Lorentz-invariance and the requirement that the equations of motion for the decomposed field $(h_{\m\n}^T,\, A_\m,\,\phi)$ are at most of second order. The results are precisely the linearized Einstein-Hilbert kinetic term plus the Fierz-Pauli mass term. 

\subsection{Two chains of antisymmetrized indices}\label{subs-chains}
In this subsection, two chains of antisymmetrized indices are constructed from the assumptions of Lorentz-invariance and ghost-freeness. 
We first discuss the derivative indices in this subsection and then the tensor indices in the next subsection \ref{subs-tenind}, 
which are parallel to subsection \ref{subs-Ost} and subsection \ref{subs-lon}. 

A Lagrangian will not lead to higher order equations of motion if it is constructed from zeroth and first order derivative terms. 
Here ``higher order'' means the equations of motion are of more than second order. 

Let us consider a Lagrangian which contains harmless zeroth order terms and dangerous second order terms
\be
\mathcal L\sim \phi\dots \phi \,\pa\pa\phi\dots \pa\pa\phi.
\ee
For the moment, we suppress the tensor indices in $\phi$, where $\phi$ could be tensor fields. 
The tensor indices are discussed in next subsection \ref{subs-tenind}. 
We also leave the discussion of first order terms to subsection \ref{subs-1order}. 

There must be zeroth order terms (or first order terms) in the Lagrangian to have equations of motion of at most second order. 
Without the zeroth order terms, 
the variation of the second order term will generate third or fourth order terms in the equations of motion, 
unless they vanish due to some cancellation and then these terms have no dynamics.  

At most second order terms are considered in the Lagrangians because of the presence of zeroth order terms. 
The variations of the zeroth order terms will generate terms of the same orders as those in the Lagrangian, 
so usually a higher order (cubic, fourth,\dots) term in the Lagrangian will lead to the same higher order term in the equations of motion 
when there are zeroth order terms, 
unless it is cancelled by other terms from the variations of high order terms. 
As we explain below, there are only two chains of antisymmetrized indices for the derivatives, 
so terms with higher order derivatives will simply vanish. 

Now let us examine the equations of motion. To capture the essence, we focus on the product of two second order terms
\be
\mathcal L=\pa_{a_1}\pa_{a_2}\phi\,\pa_{b_1}\pa_{b_2}\phi\dots
\ee
 where $\dots$ includes the zeroth order terms and other second order terms.
 
Varying the action $\int d^D x\,\mathcal L$ with respect to $\phi$ gives us the equations of motion
\ba
\frac{\pa \mathcal L }{ \pa \phi}=&&\,\pa_{a_1}\pa_{b_1}\pa_{b_2}\phi\,\pa_{a_2}(\dots)+\pa_{a_2}\pa_{b_1}\pa_{b_2}\phi\,\pa_{a_1}(\dots)
\nonumber\\&&\,
+\,\pa_{b_1}\pa_{a_1}\pa_{a_2}\phi\,\pa_{b_2}(\dots)+\pa_{b_2}\pa_{a_1}\pa_{a_2}\phi\,\pa_{b_1}(\dots)
\nonumber\\&&\,
+\dots\quad\label{dLdphi}
\ea

Let us concentrate on the first term. 
The derivative indices are contracted with a constant Lorentz-covariant tensor
\be
C^{\m,\r\s\dots}\pa_{\m}\pa_{\r}\pa_{\s}\phi\,(\dots)\,.
\ee
The derivatives commute with each others, so the coefficient of a third order term is a sum over $C$ with different orders of indices. 
For example, the explicit coefficient of a specific term is 
\ba
(C^{1,23\dots}+C^{1,32\dots}+C^{2,13\dots}
+C^{2,31\dots}
\nonumber\\
+C^{3,12\dots}+C^{3,21\dots})
\pa_{1}\pa_{2}\pa_{3}\phi\,(\dots).
\ea
To avoid the unwanted third order term, we require the covariant tensor $C$ to solve
\ba
&\,C^{\m,\r\s\dots}+C^{\m,\s\r\dots}+C^{\r,\m\s\dots}+C^{\r,\s\m\dots}\label{loc}
\nonumber\\
&\,
+C^{\s,\r\m\dots}
+C^{\s,\m\r\dots}=0\label{loc}.
\ea
This is a ``local'' equation which guarantees the coefficient of $\pa_{a_1}\pa_{b_1}\pa_{b_2}\phi$ vanishes 
by considering only the two second order terms. 
There could be theories whose $C$ do not solve this local equation, 
but the coefficient of the third order term still vanishes 
by taking into account the contribution from other terms.
However, no example of such kind is known. 

We first give an intuitive solution to eq. \eqref{loc}.  \
Assume the building blocks of $C$ are the Minkowski metric $\eta_{\m\n}$ and its inverse $\eta^{\m\n}$, 
which do not include the Levi-Civita symbol. 
We can expand $C$ in terms of $\eta$
\ba
C^{\m,\r\s\dots}=&&\,
\eta^{\m\r}\eta^{\s\a} {(C_1)_{\a}}^{\dots}
+\eta^{\m\s}\eta^{\r\a} {(C_2)_{\a}}^{\dots}
\nonumber\\
&&\,+\eta^{\r\s}\eta^{\m\a} {(C_3)_{\,\,\a}}^{\dots}\quad.\label{expand}
\ea

There could be one more term
\be
\eta^{\m\a}\eta^{\r\b}\eta^{\s\g}
{(C_4)_{\a\b\g}}^{\dots}\quad,
\ee  
but since $C_4$ are constructed from $\eta$, 
two of the three indices $(\a,\,\b,\,\g)$ will be contracted with $\eta$ 
and thus the $C_4$ term is not independent.

The equation \eqref{loc} requires
\be
(\eta^{\m\r}\eta^{\s\a} 
+\eta^{\m\s}\eta^{\r\a} 
+\eta^{\r\s}\eta^{\m\a} 
 ){(C_1+C_2+C_3)_{\a}}^{\dots}=0.\label{solu}
\ee
The solution of $C$ is
\ba
C^{\m,\r\s\dots}=&\,
(\eta^{\m\r}\eta^{\s\a} -\eta^{\s\r}\eta^{\m\a} ){(C_1)_{\a}}^{\dots}
\nonumber\\&\,
+(\eta^{\m\s}\eta^{\r\a}-\eta^{\r\s}\eta^{\m\a}) {(C_2)_{\a}}^{\dots}
\ea
which can be reduced to one term by changing the order of the indices in the second derivative. 

Therefore, the first term in eq. \eqref{dLdphi} vanishes when 
either $ (a_1,\,b_1)$ or $ (a_1,\,b_2)$ are antisymmetrized. 
The formal way to derive the same solution is decomposing $C^{\m,\r\s\dots}$ according to the permutation group. 
The totally antisymmetric part vanishes as the indices $(\r,\,\s)$ are contracted with the second derivative indices. 
The local equations \eqref{loc} indicates the totally symmetric part should vanish, 
so we are left with the mixed parts and arrive at the same result. 

For the second term in eq. \eqref{dLdphi}, we also have two choices:
either $(a_2,\,b_1)$ or $(a_2,\,b_2)$ are antisymmetrized. 
Therefore, we have two ansatzes to avoid higher order terms in the equations of motion
\begin{itemize}
\item
$(a_1,\,b_1)$ and $(a_2,\,b_2)$ are two sets of antisymmetrized indices;
\item
$(a_1,\,b_2)$ and $(a_2,\,b_1)$ are two sets of antisymmetrized indices.
\end{itemize}

Then we impose the same requirements to other combinations of second derivative terms. 
This lead us to two chains of antisymmetrized indices. 
The indices in the second order derivative must be in one of the antisymmetric chains. 
The third and higher order terms are not allowed in the Lagrangians 
because the third derivative index and one of the first two indices are antisymmetrized. 

\subsection{Tensor field indices}\label{subs-tenind}
As discussed in subsection \ref{subs-lon}, it is not sufficient to have equations of motion of at most second order 
to avoid ghost-like degrees of freedom. 
The longitudinal modes in the tensor fields could lead to higher order equations of motion 
and they are still dangerous. 
Among them, the longitudinal scalar modes are the most dangerous because all the tensor indices become derivative indices
\be
\phi_{\m_1\m_2\dots\m_s}=\pa_{\m_1}\pa_{\m_2}\dots\pa_{\m_s}\Phi.
\ee
We should require the equations of motion for the decomposed field are at most of second order. 

Firstly, the tensor indices in the second order terms should be antisymmetrized 
according to the analysis in the previous subsection \ref{subs-chains}. 
Secondly, the tensor indices of the spin-2 fields without derivative should be antisymmetrized as well 
because a decomposed spin-2 field according to eq. \eqref{spin2-de} contains a second order term of the longitudinal scalar mode. 
Thirdly, the tensor indices of the symmetric spin-2 field should be in different chains, otherwise this term will vanish due to antisymmetry. 

To antisymmetrize all the tensor field indices is a sufficient 
but not necessary condition for at most second order equations of motion for the decomposed fields. 
In principle, higher order terms could vanish without the help of the antisymmetric structure 
if two of the derivative indices are already antisymmetrized
or the derivatives are contracted with the decomposed fields 
(see eq. \eqref{trans} for example). 
However, we do not know any example where all the longitudinal higher order terms vanish without the antisymmetric structure. 

\subsection{First order terms}\label{subs-1order}
First order terms are the first derivative of the dynamical fields
\be
\pa_\m\Phi,\quad \pa_\m A_\n,\quad\pa_\m h_{\n\r}.
\ee
They are dangerous when the Lagrangian contains second order terms. 
To explain this, let us consider the product of a first order term and a second order term
\be
\mathcal L=\pa_{a_1} \phi_1\, \pa_{b_1}\pa_{b_2}\phi_2\dots
\ee
where $\phi_1$ and $\phi_2$ could be tensor fields and they could be the same or different dynamical fields. 
The equation of motion has a third order term
\be
\frac{\d \mathcal L} {\d \phi_1}=-\pa_{a_1}\pa_{b_1}\pa_{b_2}\phi_2\dots+\dots
\ee

Based on the analysis of subsections \ref{subs-chains} and \ref{subs-tenind}, 
the derivative and tensor field indices of the first order terms should be antisymmetrized. 
The tensor indices should be antisymmetrized 
because the first derivative of the tensor fields are at least of second order for the longitudinal scalar modes.  

However, there is an exception for the Lagrangians of single scalar field 
\be
\mathcal L=\pa_{a_1} \Phi\, \pa_{b_1}\pa_{b_2}\Phi\dots,
\ee
where $\Phi$ is a scalar field without tensor indices. 
The equation of motion has two third order terms
\be
\frac{\d \mathcal L} {\d \Phi}=-\pa_{a_1}\pa_{b_1}\pa_{b_2}\Phi+\pa_{b_1}\pa_{b_2}\pa_{a_1}\Phi\dots+\dots
\ee
which cancel out due to the different numbers of integrations by parts. 

In general, the exceptional cases where first order terms are harmless are: 
after decomposition the scalar field in the first order terms are the only scalar field in the second order terms. 

\subsection{Single-index terms}\label{subs-sindex}
A single-index term is a term with only one index. 
They are the first derivative of the scalar fields and the spin-1 fields without derivative
\be
\pa_\m\Phi,\quad A_\m.
\ee
The concepts of single-index terms are different from that of the first order terms which are the first derivative of dynamical fields. 

The scalar field without derivative (zero-index terms) are always harmless. 
Single-index terms could be harmless as well. 
In contrast, the other terms are usually dangerous and should be antisymmetrized
\footnote{ There are exceptional cases. 
For example, we could construct the theories from the field strengths of spin-1 fields. 
The longitudinal modes the field strengths are absent. 
These two-index terms are similar to the first derivatives of the scalar field, 
so they could be harmless even if they have two indices. }. 

Based on the analysis of subsection \ref{subs-1order}, 
$\pa_\m\Phi$ is dangerous when there are terms like. 
\ba
&&\pa_\m\pa^\n\tilde\Phi,\,
\nonumber\\
&&
\pa_\m A^\n,\,\pa^\n A_\m,\,
\pa_\m\pa^\n A_\m,\,\pa_\m\pa^\n A^\n,
\nonumber\\
&&{h_\m}^\n,\,\pa_\m {h_\m}^\n,\,\pa^\n {h_\m}^\n,\,\pa_\m\pa^\n {h_\m}^\n,
\ea
where $\tilde \Phi$ is a different scalar field. 
 
The longitudinal scalar field $\Phi$ in $A_\m$ is more dangerous than the transverse vector field
\be
A_\m=A_\m^T+\pa_\m \Phi.
\ee

$A_\m$ is dangerous when there are terms like 
\ba
&&\pa_\m\pa^\n\tilde\Phi,\,
\nonumber\\
&&
\pa_\m \tilde A^\n,\,\pa^\n \tilde A_\m,\,
\pa_\m\pa^\n \tilde A_\m,\,\pa_\m\pa^\n \tilde A^\n,
\nonumber\\
&&
\pa_\m\pa^\n A_\m,\,\pa_\m\pa^\n A^\n,\,
\nonumber\\
&&{h_\m}^\n,\,\pa_\m {h_\m}^\n,\,\pa^\n {h_\m}^\n,\,\pa_\m\pa^\n {h_\m}^\n,
\ea
where $\tilde \Phi$ is different from the longitudinal scalar mode in $A_\m$, 
$\tilde A$ is a different spin-1 field. 
The case $\pa_\m\pa^\n A_\m$ is discussed in detail in subsection \ref{subs-exs1}. 

\subsection{General ghost-free Lorentz-invariant Lagrangians}\label{subs-genL}
From the two antisymmetric chains, we can construct the antisymmetric Kronecker delta
\be
\d^{\m_1\m_2\dots\m_k}_{\n_1\n_2\dots\n_k}=\d_{\n_1}^{[\m_1}\d_{\n_2}^{\m_2}\dots \d_{\n_k}^{\m_k]},
\ee
where $[\dots]$ are unnormalized antisymmetrization. 

The antisymmetric Kronecker delta can be constructed from two Levi-Civita symbols 
and some Minkowski metrics
\ba
&&\,\d^{\m_1\m_2\dots\m_k}_{\n_1\n_2\dots\n_k}
\nonumber\\
=&&\,\frac{
\e_{\n_1\n_2\dots\n_D}\e^{\m_1\m_2\dots\m_D}\,
{(\eta\, \eta^{-1})_{\m_{k+1}}}^{\n_{k+1}}\dots{(\eta\, \eta^{-1})_{\m_{D}}}^{\n_{D}}} {{k!(D-k)!}}.
\nonumber\\
\label{2Levi}
\ea

The general ghost-free Lagrangians are
\be
\mathcal L=\sum d^Dx \, f\, \d^{\m_1\m_2\dots}_{\n_1\n_2\dots}\prod {\o_{\m\dots}}^{\n\dots},\label{gf-Lag}
\ee
where $f$ are scalar functions of the harmless terms. 
To simplify the notation, the subscripts $i$ in the indices $\m_i$ and $\n_i$ are not written explicitly in $\o$. 

The scalar functions $f$ could be functions of the scalar field $\Phi$. 
If some single-index terms are harmless according to discussion of subsection \ref{subs-sindex}, 
they can live in the scalar function $f$ as well. 

In the products, each $\o$ could be different from the others and they could be chosen from
\ba
&&\,\pa_\m\Phi,\quad\pa^\n\Phi,\quad\pa_\m\pa^\n\Phi,
\nonumber\\
&&\,A_\m,\quad\pa_\m A_\m,\quad\pa^\n A_\m,\quad\pa_\m\pa^\n A_\m,
\nonumber\\
&&\,A^\n,\quad\pa_\m A^\n,\quad\pa^\n A^\n,\quad\pa_\m\pa^\n A^\n,
\nonumber\\
&&\,{h_\m}^\n,\quad\pa_\m {h_\m}^\n,\quad\pa^\n {h_\m}^\n,\quad\pa_\m\pa^\n {h_\m}^\n.\label{o}
\ea

The ghost-free terms are the products of scalar functions $f$ and the building blocks $\o$ in \eqref{o}, 
where the indices of the building blocks are antisymmetrized 
using the antisymmetric Kronecker delta. 

By construction, the equations of motion for the decomposed fields are at most of second order 
thanks to the antisymmetric structure. 

Below are a few comments on the general Lagrangians:
\begin{itemize}
\item
To derive the antisymmetric Kronecker delta, one chain of antisymmetrized indices are raised by the Minkowski metric. 
It is assumed that the lengths of the two chains are the same and the indices are contracted with each others. 
There might be exceptions and some of the indices in one chain is not contracted with the indices in the other chain. 
These free indices are then contracted with the harmless single-index terms, 
which are the only possible terms that are not antisymmetrized. 
Due to the antisymmetric structure of the chain indices, 
the indices of the harmless single-index terms will be absorbed to the other chain. 
The Lagrangians are still in the form of eq. \eqref{gf-Lag}.
\item
To ensure the Hamiltonian is bounded from below, the signs of the scalar functions $f$ are vary important. 
It is possible that the energy is bounded from above due to a wrong sign of $f$ 
in spite of the absence of higher order terms in the equations of motion. 

\item
By field redefinitions, integration by parts and using the properties of the antisymmetric Kronecker delta, 
a ghost-free Lagrangian can be transformed into a different form. 
If the new formulation again takes the form of eq. \eqref{gf-Lag}, the two Lagrangians are equivalent. 
In most of the cases, the new formulation is not in the form of eq. \eqref{gf-Lag}, then one way to show the new Lagrangian is ghost-free is undoing the changes and then bringing it back to the antisymmetric form. 

For example, the $f(R)$ gravity seems to be propagating ghost-like degrees of freedom 
because the second order derivative indices are not totally antisymmetrized. 
However, by field  redefinitions, a $f(R)$ theory is equivalent to a scalar-tensor theory which does not contain dangerous high order terms. 
One can conclude that no ghost-like degrees of freedom is generated by high order terms in the $f(R)$ gravity. 
\item
There could be nonlinear extensions for the general ghost-free Lagrangians \eqref{gf-Lag}. 
This is very common for the theories of spin-2 fields. 
For example, the Einstein-Hilbert action is an infinite series in terms of the metric perturbation $h_{\m\n}$ 
and only the leading term takes an antisymmetric form. 
In subsection \ref{subs-exs2}, the nonlinear completion for the ghost-free theories of spin-2 field are investigated. 

In general, the lowest order terms of the nonlinear ghost-free Lagrangians should take the form of eq. \eqref{gf-Lag}. 
This is because in the weak field region, the Lagrangians are dominated by the lowest order terms 
and the higher order terms can be neglected. 
The theories should be still free of ghost-like degrees of freedom. 
Without the complication from the infinite numbers of nonlinear terms, 
the expressions of the lowest order terms can be determined to be eq. \eqref{gf-Lag}. 
\end{itemize}

\section{The framework in the language of differential forms}\label{sec-form}
In the section, the general framework are reformulated in the language of differential forms. 
In subsection \ref{subs-moti}, the motivations to introduce differential forms are explained. 
Then the general Lagrangians are translated into the language of differential forms in subsection \ref{subs-tran}. 
In this formulation, the absence of higher order terms originates in a basic fact in differential geometry, 
which is discussed in subsection \ref{subs-natgf}. 
In the subsection \ref{subs-dua}, a novel duality of the general ghost-free Lagrangians is proposed, 
which generalizes the Hodge duality in Maxwell's theory. 
A novel mathematical concept, ``double form'', is proposed in subsection \ref{subs-douform}. 

\subsection{Motivations}\label{subs-moti}
The first motivation to use the language of differential forms is from 
the crucial importance of antisymmetrization in the general ghost-free Lagrangians \eqref{gf-Lag}. 
It is natural to replace the antisymmetrization among the indices of different second order terms 
with an antisymmetric product. 
The wedge product in differential geometry is a natural choice.

The second motivation comes from the fact that differential form provides a unified approach 
to defining the integrands. 
The Lagrangians are the integrands of the action which integrate over the spacetime manifolds, 
such as curves, surface, volumes. 
As a unifying framework, the general ghost-free Lagrangians should be reformulated as differential forms.

Usually, the Lagrangians are some trivial products of scalar functions and the volume forms $d^Dx$, 
where the scalar functions are known as Lagrangian densities. 
In contrast, for high derivative or tensor field theories, the Lagrangians have non-trivial structures. 
For example, the Lovelock terms in the language of differential forms are the wedge product of 
curvature two-forms and vielbeins
\be
\mathcal L_{\text{Lovelock}}\sim R\wedge \dots \wedge R\wedge E\wedge\dots\wedge E,\label{LL-df}
\ee
where $R$ is the curvature two-form and $E$ is the dynamical vielbein. 
We can see the curvature two-form is part of the differential forms in the wedge products. 
This leads us to the next motivation. 

The third motivation is due to the Zumino's elegant reformulation \cite{Zumino:1985dp}  of the Lovelock theories \eqref{LL-df}. 
The Lovelock gravity is the most general metric theory that leads to second order equations of motion. 
Although the Lovelock terms \eqref{Lovelock} are the products of the Riemann curvature tensor, which are second order terms, 
the equations of motion do not contain higher order terms. 

In the formulation of differential forms, the equations of motion are obtained by varying the integrals of the Lovelock terms \eqref{LL-df}. 
The variations of the curvature two-forms give the covariant derivative of the variations of spin connection
\be
\d R=D (\d \o),
\ee
where $\o$ is the spin connection. 
After integrating by parts, the covariant derivative acts on other curvature two-forms. 
They are third order terms in the equations of motion. 
However, it is known that these dangerous terms vanish thanks to the Bianchi identities
\be
D R=0,
\ee
which originates in the basic fact that the square of exterior derivative vanishes
\be
d^2=0. 
\ee
The Lovelock theory is ghost-free because every exact form is closed or the boundary of a boundary vanishes. 
The third motivation for introducing the language of differential forms is 
to generalize this statement to other ghost-free theories.  

\subsection{Translation}\label{subs-tran}
It is straightforward to translate the general ghost-free Lagrangians 
from the language of tensors into that of differential forms:
\begin{enumerate}
\item
Substitute the volume form $d^Dx$ and the $\m$-Levi-Civita symbol 
with the wedge products of $dx^{\m_i}$
\be
d^Dx\,\e^{\m_1\m_2\dots\m_D}\rightarrow dx^{\m_1}\wedge dx^{\m_2}\wedge\dots\wedge dx^{\m_D};
\ee
\item
Construct the Minkowski vielbein
\be
{\eta_\m}^\n dx^\m\rightarrow \eta^\n;
\ee
\item
Introduce more differential forms
\be
{\o_{\m_1\dots}}^{\n_1\dots}dx^{\m_1}\dots\rightarrow \o^{\n_1\dots}\,. \label{o-form}
\ee
where ${\o_{\m_1\dots}}^{\n_1\dots}$ are chosen from \eqref{o}.
\end{enumerate}
The translation is completed after these three steps. 
To illustrate the abstract procedures of translation, we consider the case of single scalar field in Appendix \ref{app-form}, 
where the translation is worked out step by step in an explicit way. 

Therefore, the general ghost-free Lagrangians are D-forms
\be
\mathcal L=\sum\,f\,\e_{\n_1\dots\n_D}\,\o^{\n_1}\wedge\dots\wedge\o^{\n_k}\wedge \eta^{\n_{k+1}}\wedge\dots\wedge\eta^{\n_D},
\ee
where $\o$ could be different differential forms. 

Schematically, the general ghost-free Lagrangians read
\be
\mathcal L=\sum\,f\,\o\wedge\dots\wedge\o, \label{ge-df}
\ee
where $\o$ could be different matter differential forms and geometric differential forms. 
The matter forms are built from \eqref{o} and eq. \eqref{o-form}. 
The geometric forms are nonlinear, geometric completion for the differential forms constructed from the fourth line of \eqref{o}. 
$\eta$ is considered as a geometric form as well. 
The nonlinear completion of the geometric forms are discussed in subsection \ref{subs-exs2}.

In the above translation, we use the $\m$-Levi-Civita symbol to construct differential forms. 
Since there are two Levi-Civita symbols, we could use the $\n$-Levi-Civita symbol instead. 
Therefore, the general ghost-free Lagrangians have dual formulations by exchanging the $\m_i$ and $\n_i$ indices. 
This duality is discussed in more detail in subsection \ref{subs-dua}. 

\subsection{Geometric nature of ghost-freeness}\label{subs-natgf}
Now let us examine how the higher order terms are avoided in the formulation of differential forms. 
By construction, the derivative terms are exact forms. 
For example, the second order terms are
\be
\pa_\m\pa^\n\phi\, dx^\m=d(\pa^\n \phi),
\ee
where $\phi$ could be tensor fields. 
For simplicity, the tensor indices of $\phi$ and the corresponding $dx^\m$ are not written explicitly.

Varying the action with respect to the dynamical field, the exterior derivative will move to other terms. 
If the exterior derivative acts on a second order term, third order terms can be generated in the equations of motion. 
However, the second order terms are exact forms and their exterior derivatives vanish due to
\be
d^2=0 
\ee 
and the higher order terms will not appear in the equations of motion. 

Therefore, the statement concerning the nature of the ghost-freeness in Lovelock's gravity in subsection \ref{subs-moti} is generalized to other ghost-free theories in the general framework. 

The general statement is that 
the dangerous higher order terms are absent because every exact form is closed. 
This is the geometric nature of ghost-freeness, which follows from the ``local'' ghost-free equation \eqref{loc}. 

To complete the discussion of higher order terms, one should go to the dual formulation 
which concerns with the $\n$-derivatives. 
Higher order terms induced by additional $\n$-derivatives are also absent because of the same property of exterior derivative.

\subsection{Novel duality}\label{subs-dua}
In four dimensions, the vacuum Maxwell's equations are invariant under the electromagnetic duality. 
In the language of differential forms, the Maxwell kinetic term
\be
\mathcal L_{\text{Maxwell}}\sim F\wedge\ast F
\ee
 is invariant under the Hodge star operator
 \be
 F\rightarrow \ast F.
 \ee

In the general framework, the Maxwell kinetic term in the form of \eqref{gf-Lag} reads
\be
\mathcal L_{\text{Maxwell}}\sim d^Dx\,\d_{\n_1\n_2}^{\m_1\m_2}
\,
\pa_{\m_1}A_{\m_2}\pa^{\n_1}A^{\n_2}. 
\ee
The Hodge star operator is a transformation that exchanges $\m_i$ and $\n_i$ indices
\be
\m_i\leftrightarrow \n_i,  
\ee  
which is the new duality discussed at the end of subsection \ref{subs-tran}. 
This novel duality for general ghost-free theories is a generalization of the Hodge duality in Maxwell's theory. 

Since the dual formulations are the same, the Maxwell kinetic term is self-dual under this duality transformation.  
Other examples of self-dual Lagrangians are the Galileons, the dRGT terms (in the symmetric gauge for the dynamical vielbeins), and the Lovelock terms (which include the Einstein-Hilbert term). 
The self-duality of the Lovelock terms is more transparent in the formulation of double forms 
discussed in the next subsection \ref{subs-douform}.  

\subsection{Double forms}\label{subs-douform}
The novel duality comes from the ambiguity in choosing one of the Levi-Civita symbols to construct differential forms. 
Here we propose a natural mathematical concept that is related to this ambiguity. 
We call it ``double form'' as both of the two Levi-Civita symbols are used to construct differential forms. 

As double forms, the general ghost-free Lagrangians are $(D,\,D)$-forms. 
The reformulations of the Maxwell kinetic term and the Einstein-Hilbert term 
in terms of double forms 
and the self-duality of the Lovelock terms are discussed in Appendix \ref{app-douf}. 

From the perspective of action, the additional D-form seems to be redundant 
as an action is an integral over spacetime manifold with the volume form being a D-form. 
Here we think of the actions as formal devices for arriving at the equations of motion. 

We want to emphasis that the use of double-form makes some properties of the ghost-free Lagrangian more transparent:
\begin{itemize}
\item 
A duality transformation is an exchange of the two D-forms in a Lagrangian. 
The self-duality of the Lovelock terms is clear only in the double form formulations.
\item
The higher order terms are absent in the equations of motion because a second order term is a double exact-form and 
it vanishes under the action of either $\m$ or $\n$ exterior derivative.
\end{itemize}

There is one more important motivation for proposing ``double form''.
It comes from the general ghost-free Lagrangians of p-form fields, 
which is discussed in subsection \ref{subs-expf}. 
In this case, the concept of ``multiple form" is unavoidable.  

\section{Examples}\label{sec-ex}
In this section, some concrete examples of the general Lagrangians are discussed. 
We start from the simplest case, the theories of spin-0 fields, in subsection \ref{subs-exs0}. 
Then we move to the theories of spin-1 fields in subsection \ref{subs-exs1}. 
There are two main directions in extending the discussions of tensor fields. 
The first direction is to consider the antisymmetric extension, p-form fields, which is studied in subsection \ref{subs-expf}. 
The second direction is to study the symmetric extension, symmetric spin-2 fields, which are related to gravity. 
This is discussed in subsection \ref{subs-exs2}, where 
nonlinear completion for the spin-2 theories are investigated as well. 
We will not discuss the mixed extensions of tensor fields. 
In the end, the coupling among different kinds of dynamical fields are discussed in subsection \ref{subs-coupled}.

\subsection{Spin-0 fields}\label{subs-exs0}
Using the building blocks in the first line of \eqref{o}, the general ghost-free Lagrangians for scalar fields are
\be
\mathcal L=\sum\,d^Dx\, f\,\d_{\n_1\dots}^{\m_1\dots}[\pa_\m\Phi][\pa^\n\Phi][\pa_\m\pa^\n\Phi],
\label{ge-s0}
\ee
where $[X]$ are products of $X$ 
\be
[X]=\prod X
\ee
and they could simply be $1$. 
The subscripts $i$ of $\m_i,\,\n_i$ in the products are not written explicitly. 
$\Phi$ could be different scalar fields. 
The scalar functions $f$ could depend on the scalar fields and their first derivatives. 

\paragraph{Single scalar field}
In the case of single field, we have two sets of possible terms in the ghost-free Lagrangians
\ba
\mathcal L_1^{(k)}
=&&\,d^Dx \,f\, \d_{\n_1\dots }^{\m_1\dots }
[\pa_{\m}\pa^{\n}\Phi]
\nonumber\\
=&&\,
d^Dx \,f\, \d_{\n_1\dots \n_k}^{\m_1\dots \m_k}
\prod_{i=1}^k \pa_{\m_i}\pa^{\n_i}\Phi\label{sc1}
\ea
and
\ba
\mathcal L_2^{(k)}=&&\,d^Dx \,f\, \d_{\n_1\dots }^{\m_1\dots }(\pa_\m\Phi)(\pa^\n\Phi)
[\pa_{\m}\pa^{\n}\Phi]
\nonumber\\
=&&\,d^Dx \,f\, \d_{\n_1\dots \n_k}^{\m_1\dots \m_k}\,
(\pa_{\m_1}\Phi\,\pa^{\n_1}\Phi)
\prod_{i=2}^k \pa_{\m_i}\pa^{\n_i}\Phi,
\label{sc2}
\ea
where the scalar functions are functions of $\Phi$ and $\pa_\r\Phi\,\pa^\r\Phi$
\be
f=f(\Phi,\, \pa_\r\Phi\,\pa^\r\Phi).
\ee

Eq. \eqref{sc1} has no single-index term, while eq. \eqref{sc2} contains two single-index terms.  

According to the discussion of subsection \ref{subs-sindex}, $f$ could depend on $\pa_\r\Phi$.
Since $f$ are scalar functions, the free index of $\pa_\r\Phi$ should be contracted with another $\pa_\r\Phi$. 
Therefore, $f$ depend on $\Phi$ and $\pa_\r\Phi\,\pa^\r\Phi$. 

These two Lagrangians are related by integrating by parts and using the properties of the antisymmetric structure. 
They are precisely two equivalent formulations of Galileons, 
which are the most general single scalar theories 
that are Lorentz-invariant and lead to at most second order equations of motion. 

In the case of single scalar field, we can not construct new theories due to the small numbers of indices. 
Below we consider two scalar fields and there are considerably richer possibilities. 

\paragraph{Bi-scalar fields}
For two scalar fields, we have five sets of possible terms in the ghost-free Lagrangians
\be
\mathcal L_1=d^Dx\, f\,\d_{\n_1\dots}^{\m_1\dots}[\pa_\m\pa^\n\Phi_1][\pa_\m\pa^\n\Phi_2],
\qquad\qquad\qquad
\label{biG}
\ee
\be
\mathcal L_2=d^Dx\, f\,\d_{\n_1\dots}^{\m_1\dots}(\pa_\m\Phi_1)\,(\pa^\n\Phi_1)\,[\pa_\m\pa^\n\Phi_1][\pa_\m\pa^\n\Phi_2],
\ee
\be
\mathcal L_3=d^Dx\, f\,\d_{\n_1\dots}^{\m_1\dots}(\pa_\m\Phi_2)\,(\pa^\n\Phi_2)\,[\pa_\m\pa^\n\Phi_1][\pa_\m\pa^\n\Phi_2],
\ee
\be
\mathcal L_4=d^Dx\, f\,\d_{\n_1\dots}^{\m_1\dots}(\pa_\m\Phi_1)\,(\pa^\n\Phi_2)\,[\pa_\m\pa^\n\Phi_1][\pa_\m\pa^\n\Phi_2],
\ee

\ba
\mathcal L_5=d^Dx\, f\,\d_{\n_1\dots}^{\m_1\dots}\,(\pa_\m\Phi_1)\,(\pa^\n\Phi_1)\,(\pa_\m\Phi_2)\,
(\pa^\n{\Phi_2})\qquad\quad
\nonumber\\\
[\pa_\m\pa^\n\Phi_1][\pa_\m\pa^\n\Phi_2],\quad
\ea
where $f$ are functions of $\Phi_1$ and $\Phi_2$
\be
f=f(\Phi_1,\,\Phi_2).
\ee
Based on the discussion of subsection \ref{subs-sindex}, 
if a term does not contain $\pa\pa\Phi_2$, 
then $f$ could depend on $\pa_\r\Phi_1\pa^\r\Phi_1$. 
In parallel, if a term does not have $\pa\pa\Phi_1$, then $f$ could be a function of $\pa_\r\Phi_2\pa^\r\Phi_2$. 
These two cases are similar to the theories of single scalar field. 
The bi-galileons \cite{Padilla:2010de} are special cases of eq. \eqref{biG} where 
\be
f =a_1 \Phi_1+a_2\Phi_2.
\ee

Besides bi-Galileons, multi-Galileons were investigated using the brane-world constructions \cite{Hinterbichler:2010xn}. 
Their structure are similar to the bi-Galileons and they are special cases of eq. \eqref{ge-s0} as expected. 

\subsection{Spin-1 fields}\label{subs-exs1}
Using the building blocks in the second and third lines of \eqref{o}, the general ghost-free Lagrangians for vector fields read
\ba
\mathcal L=\sum\,d^Dx\, f\,\d_{\n_1\dots}^{\m_1\dots}
[ A_\m][ A^\n]
[\pa_\m A_\m][\pa^\n A_\m]
\qquad\nonumber\\\
[\pa_\m A^\n][\pa^\n A^\n]
[\pa_\m\pa^\n A_\m]
[\pa_\m\pa^\n A^\n]\label{ge-s1},
\ea
where $A_\m$ could be different spin-1 fields and $f$ could depend on $A_\m A^\m$. The notations are explained 
below eq. \eqref{ge-s0}.

For single spin-1 field, there are three sets of ghost-free terms
\ba
\mathcal L_1=\sum\,d^Dx\, f\,\d_{\n_1\dots}^{\m_1\dots}
[\pa_\m A_\m][\pa^\n A_\m]
\qquad\nonumber\\\
[\pa_\m A^\n][\pa^\n A^\n]
[\pa_\m\pa^\n A_\m]
[\pa_\m\pa^\n A^\n],
\label{s1-L1}
\ea

\ba
\mathcal L_2=\sum\,d^Dx\, f\,\d_{\n_1\dots}^{\m_1\dots}
[ A_\m]
[\pa_\m A_\m][\pa^\n A_\m]
\qquad\nonumber\\\
[\pa_\m A^\n][\pa^\n A^\n]
[\pa_\m\pa^\n A_\m]
[\pa_\m\pa^\n A^\n],\label{s1-L2}
\ea
\ba
\mathcal L_3=\sum\,d^Dx\, f\,\d_{\n_1\dots}^{\m_1\dots}
[ A_\m][ A^\n]
[\pa_\m A_\m][\pa^\n A_\m]
\qquad\nonumber\\\
[\pa_\m A^\n][\pa^\n A^\n]
[\pa_\m\pa^\n A_\m]
[\pa_\m\pa^\n A^\n],\label{s1-L3}
\ea
where $\mathcal L_1$ has no single-index term in the antisymmetric products, 
$\mathcal L_2$ and $\mathcal L_3$ contain one and two single-index term in the products.

If second order terms $\pa\pa A$ exist, $f$ can only be constants. 
Otherwise, $f$ could depend on $A_\r A^\r$. 
In other words, when there are second order terms, 
the indices of the single-index terms $A_\m$ should be antisymmetrized even though we are discussing single field theories. 
This is based on the analysis of subsection \ref{subs-sindex}. 
Below we explain this point in more detail. 

After one decomposes the spin-1 field into the transverse and the longitudinal parts
\be
A_\m=A_\m^T+\pa_\m \Phi, \quad \pa^\m A_\m^T=0,
\ee
the second order terms become
\be
\pa_\m\pa^\n A_\m\rightarrow \pa_\m\pa^\n A_\m^T,\quad
\pa_\m\pa^\n A^\n\rightarrow \pa_\m\pa^\n (A^T)^\n.
\ee
For example, when the Lagrangians are constructed 
from $ A_{\m_1}$ and $\pa_{\m_2}\pa^\n A_{\m_3}$, we have a dangerous term
\be
\pa_{\m_1} \Phi \,\pa_{\m_2}\pa^\n A_{\m_3}^T,
\ee
whose equations of motion contain third order terms. 
To avoid the third order term, the indices $(\m_1,\,\m_2,\,\m_3)$ should be antisymmetrized,
so the scalar function $f$ can not depend on $A_\m$.  

For U(1) gauge invariant theories, only the gauge-invariant building blocks
\be
\pa_\m A_\m,\quad\pa^\n A^\n,\quad
\pa_\m\pa^\n A_\m,\quad\pa_\m\pa^\n A^\n
\ee
are allowed. 
It was showed in \cite{Deffayet:2013tca} that, 
to avoid higher order terms in the equations of motion, 
the indices of $A_\m$ should be antisymmetrized. 
These Lagrangians are built from the special cases of \eqref{s1-L1} where $\pa_\m A^\n,\, \pa^\n A_\m$ are absent and $f$ are constants. 
However, the corresponding equations of motion vanish due to antisymmetry, 
except Maxwell's equations from the Maxwell kinetic term. 

Intuitively, the vanishing of equations of motion follows from the absence of zeroth order terms (modulo integrating by parts). 
One can consider non-trivial Lagrangian by introducing zeroth order terms. 
It is known that a zeroth order term can break the U(1) gauge symmetry. 
Therefore, the goal becomes constructing Galileon-like theories of spin-1 field without gauge invariance \cite{vector-galileon} . 
They are some special cases of eq. \eqref{ge-s1} 
without the zeroth and the second order terms in the antisymmetric products. 
They are the generalization of Proca's theory. 

The Lagrangian of Proca's theory consists of a Maxwell kinetic term and a mass term which break the gauge symmetry. 
The equations of motion for the decomposed fields $(A^T_\m,\,\Phi)$ are of second order and 
no ghost-like degrees of freedom are propagating in the Proca theory. 

The absence of ghost-like degrees of freedom can be seen from the Hamiltonian analysis as well. 
In Maxwell's theory, there is a primary constraint and a secondary constraint, which are first class constraints and eliminate $2\times 2$ degrees of freedom in phase space. 
In Proca's theory, gauge symmetry is broken by the mass term. 
The first class constraints in Maxwell's theory become second class constraints 
and their Poisson brackets on the constraint surface are proportional to the squared mass. 
Therefore, $2\times 1$ degrees of freedom are eliminated and the Proca theory have $3\times 2$ dynamical degrees of freedom. 
The longitudinal scalar ghost is killed by the second-class constraints. 

Gauge symmetries are very efficient in eliminating ghost-like degrees of freedom, 
but they are not the necessary ingredients for ghost-free theories. 
As in the Proca theory, ghost-like degrees are not propagating due to the existence of second class constraints, 
which are not related to gauge symmetries.
They eliminate less degrees of freedom than the first class constraints, 
but they can still kill the ghost-like degrees of freedom. 

One of the motivations of this work is to relax the guiding principle from gauge-invariance to ghost-freeness. 
Then we examine how the general structure of the consistent theories are constrained by this weaker assumption. 

By construction, each Lagrangian in the form of eq. \eqref{ge-s1} has at least one primary constraint for each spin-1 fields. 
Let us consider single spin-1 field for simplicity. 
The primary constraint is the zero component of the conjugate momentum
\be
\pi^0=\frac {\pa\mathcal L}{\pa \dot A_0}=F(A_0,\,\,A_i, \,\pa_i A_0,\,\,\pa_i A_j,\, \d^{ij}),\label{pi0}
\ee
whose right hand side does not contain time derivative. 
This is because the derivative indices in \eqref{ge-s1} are antisymmetrized. 
There are only two antisymmetric chains and the two $0$ indices are already used in $\pa_0 A_0$, 
so the right hand side of \eqref{pi0} do not contain any time derivative. 

The primary constraint is accompanied by a secondary constraint which preserves the primary constraint in time, 
so at least two degrees of freedom are eliminated. 
Therefore, the theories with Lagrangian \eqref{ge-s1} contain 
at most $3\times 2$ dynamical degrees of freedom and 
the longitudinal scalar ghost is eliminated, 
which is similar to the Proca theory. 
This is a model-independent statement. 

However, there could be more primary constraints as well as 
tertiary, quaternary, \dots constraints from the universal primary constraint $\pi^0=F$. 
The answers to how many degrees of freedom are indeed propagating dynamically are model-dependent.

\subsection{p-form fields}\label{subs-expf}
Scalar and vector fields can be regarded as 0-form and 1-form field. 
One direction of extending the discussions is to consider p-form fields. 
The results are straightforward generalization of those of the spin-1 fields. 

The ghost-free terms for p-form fields can be obtained from eq. \eqref{ge-s1}
by the following substitutions
\be
A_\m\rightarrow B_{\m_1\m_2\dots\m_p},\quad
A^\n\rightarrow B^{\n_1\n_2\dots\n_p},
\ee
where the indices in $B_{\m_1\m_2\dots\m_p}$ are antisymmetrized
\be
B_{\m_1\m_2\dots\m_p}=B_{[\m_1\m_2\dots\m_p]}/p!.
\ee

The general ghost-free Lagrangians for the p-form fields read
\ba
\mathcal L=\sum\,d^Dx\, f\,\d_{\n_1\dots}^{\m_1\dots}
[ B_{\m\dots}][ B^{\n\dots}]
[\pa_\m B_{\m\dots}][\pa^\n B_{\m\dots}]
\nonumber\\\
[\pa_\m B^{\n\dots}][\pa^\n B^{\n\dots}]
[\pa_\m\pa^\n B_{\m\dots}]
[\pa_\m\pa^\n B^{\n\dots}]
\label{ge-pf}.
\ea
$f$ can depend on $B$ if similar requirements are satisfied as the cases of spin-1 fields. 

In contrast to the symmetric tensor fields, 
p-form fields can be decomposed only one time despite the large number of indices. 
The decomposed fields in the p-form fields are the exact part and the non-exact part
\be
B_{\m_1\dots\m_p}=B^T_{\m_1\dots\m_p}+ \pa_{[\m_p}C_{\m_1\dots\m_{p-1}]}
\ee
or 
\ba
&&\,
B_{\m_1\dots\m_p}dx^{\m_1}\wedge\dots\wedge dx^{\m_p}
\nonumber\\
=&&\,
B^{T}_{\m_1\dots\m_p}dx^{\m_1}\wedge\dots\wedge dx^{\m_p}
\nonumber\\
&&\,+d\,(C_{\m_1\dots\m_{p-1}}dx^{\m_1}\wedge\dots\wedge dx^{\m_{p-1}}),
\ea
\be
d(d\,(C_{\m_1\dots\m_{p-1}}dx^{\m_1}\wedge\dots\wedge dx^{\m_{p-1}}))=0,
\ee
where $d$ is exterior derivative and 
here the transverse part is the non-exact part of the p-form fields. 
By construction, the equations of motion for $(B^T,\, C)$ are at most of second order. 

The p-form Galileons proposed in \cite{Deffayet:2010zh} are built form the field strength $F$
\be
F_{\m_1\dots \m_{p+1}}=\pa_{[\m_{p+1}} B_{\m_1\dots\m_p]},
\ee
so they are special cases of \eqref{ge-pf} without
\be 
B_{\m\dots},\quad B^{\n\dots},\quad
\pa_\m B^{\n\dots},\quad 
\pa^\n B_{\m\dots}. 
\ee
 
Since we have two chains of antisymmetrized indices, 
some of the form indices could coincide and
the total number of form indices could be larger than the spacetime dimension $D$. 
In addition, a true p-form field is not simply its components 
and the indices should be contracted with the wedge products of $dx$
\be
B_{\m_1\dots\m_p}\rightarrow B_{\m_1\dots\m_p}dx^{\m_1}\wedge\dots\wedge dx^{\m_p}.
\ee
\be
B^{\n_1\dots\n_p}\rightarrow B_{\n_1\dots\n_p}dx^{\n_1}\wedge\dots\wedge dx^{\n_p}.
\ee
This is the practical reason that we need the notion of ``double form'', 
as there are two chains of antisymmetrized indices. 
If we do not use the volume form to build the p-form fields, 
the Lagrangians will become ``triple forms''. 

\subsection{Spin-2 fields}\label{subs-exs2}
The general ghost-free Lagrangians for symmetric spin-2 fields are constructed from the building blocks in the fourth lines of \eqref{o}
\ba
\mathcal L=\sum\,d^Dx\,f\,\d_{\n_1\dots}^{\m_1\dots}
[ {h_\m}^\n]
[\pa_\m {h_\m}^\n][\pa^\n {h_\m}^\n]
[\pa_\m\pa^\n {h_\m}^\n],\quad\label{ge-s2}
\ea
where ${h_\m}^\n$ could be different spin-2 fields. 
$f$ are constants because the spin-2 fields themselves have already two indices 
while only single-index terms could be harmless and live in $f$. 

\paragraph{Single spin-2 field}
For the case of single spin-2 field, eq. \eqref{ge-s2} can be simplified by integrating by parts
\ba
\mathcal L=\sum\,d^Dx\,f\,\d_{\n_1\dots}^{\m_1\dots}
[ {h_\m}^\n][\pa_\m\pa^\n {h_\m}^\n].
\label{ge-s2-2}
\ea
or
\ba
\mathcal L=\sum\,d^Dx\,f_{i,j}\,\mathcal L_{i,j},
\label{ge-s2-3}
\ea
\be
\mathcal L_{i,j}=\d_{\n_1\dots\n_{2i+1}}^{\m_1\dots\m_{2i+1}}
\left(\prod_{k=1}^{i}\pa_\m\pa^\n {h_\m}^\n\right)
\left(\prod_{k=2i+1}^{2i+j} {h_\m}^\n\right),
\ee
where $i$ and $j$ indicate the numbers of $\pa\pa h$ and $h$ in the products. 

Let us see what $\mathcal L_{i,j}$ are. 
When there are no zeroth order term, $\mathcal L_{i,j}$ are total derivatives due to antisymmetrization. 
When $j=1$, $\mathcal L_{i,1}$ correspond to the non-vanishing leading perturbative terms of the Lovelock terms. 
In particular, $\mathcal L_{1,1}$ is the linearized Einstein-Hilbert term. 
If $i=0$, $\mathcal L_{0,j}$ are the perturbative terms form the ghost-free potentials. 
For example, $\mathcal L_{0,2}$ is the Fierz-Pauli term. 
$\mathcal L_{0,j}$ with $j>2$ are perturbative terms from the dRGT terms. 
$\mathcal L_{i,j}$ were studied before in \cite{Hinterbichler:2013eza} . 

The Lagrangians \eqref{ge-s2-2} are not very satisfactory from the perspective of gravitational theories. 
The actions of gravity should be a functional of the metric fields
\be
g_{\m\n}=\eta_{\m\n}+h_{\m\n},
\ee
rather than its perturbation around the Minkowski metric. 
In addition, $\mathcal L_{i,j}$ have no geometric meanings as they are just perturbative terms. 
We also expect an infinite series in terms of $h_{\m\n}$. 
Therefore, we look for nonlinear, geometric completion for \eqref{ge-s2-2}.  

We can learn a lot from the known nonlinear completion. 
The dRGT terms are nonlinear completion for $\mathcal L_{0,j}$. 
They are highly nonlinear in the language of metric, 
which make use of the square root of metric products. 
However, they are very elegant in the vielbein formulation \cite{Hinterbichler:2012cn} . 
They are simply the wedge products of the dynamical vielbein and the background vielbein
\ba
\mathcal L_{\text{dRGT}}
=&&\, E\wedge\dots\wedge E\wedge \eta \wedge\dots \wedge \eta
\nonumber\\
=&&\,\e_{\n_1\dots\n_D}E^{\n_1}\,
\wedge \dots \wedge E^{\n_k}\wedge \eta^{\n_{k+1}}\dots \wedge \eta^{\n_D},\quad
\ea 
 where $E$ is the dynamical vielbein and $\eta$ is the (Minkowski) background vielbein.
We also know the nonlinear completion for $\mathcal L_{i,0}$. They are the Lovelock terms
\ba
\mathcal L_{\text{Lovelock}}
=&&\,R\wedge \dots\wedge R \wedge E \wedge \dots \wedge E
\nonumber\\
=&&\,\e_{\n_1\dots\n_D} R^{\n_1\n_2}\,
\wedge \dots \wedge R^{\n_{2k-1}\n_{2k}}\wedge 
\nonumber\\
&&\,E^{\n_{2k+1}}\wedge\dots \wedge E^{\n_D}.
\ea

Therefore, we find a bonus of reformulating the general framework in the language of differential forms. 
It provides us very natural candidates for the nonlinear completion 
of the zeroth and the second order terms: 
they should be the dynamical vielbein and the curvature two-form! 

The natural nonlinear completion for $\mathcal L_{i,j}$ are
\be
\mathcal L=R\wedge\dots\wedge R \wedge E \wedge \dots \wedge E \wedge \eta \wedge \dots \wedge \eta,\label{nl-gr}
\ee
where $R$ is the curvature two-form of the dynamical vielbein $E$ and 
$\eta$ is the Minkowski vielbein. 
These terms are all the possible wedge products of
curvature two-form, dynamical vielbein and the background vielbein. 
The natural nonlinear completion for the first order terms are spin-connections, 
but it is less clear how to contract their indices. 

Now we have some interesting theories from Lagrangians \eqref{nl-gr}. 
In four dimensions and around Minkowski spacetime, there are two more kinetic terms for graviton
\be
\mathcal L_1=R(E)\wedge E\wedge \eta,\quad
\mathcal L_2=R(E)\wedge \eta\wedge \eta,\label{gr-kin}
\ee
where $E$ is the only dynamical vielbein, 
$R$ is the curvature two-form whose spin-connection is compatible with $E$ 
and $\eta$ is the Minkowski vielbein. 

For symmetric vielbein, the kinetic terms in \eqref{gr-kin} are metric theories in disguise. 
The kinetic terms in \cite{deRham:2013tfa}, 
which were obtained from dimensional deconstruction, 
are linear combinations of $\mathcal L_1,\,\mathcal L_2$ in \eqref{gr-kin} and the Einstein-Hilbert term. 
The minisuperspace analysis in \cite{deRham:2013tfa} showed 
the existence of abnormal $N^{-2}$  terms, 
which signals the presence of BD ghost in these kinetic terms. 
In \cite{vielbein-no-go}, negative results were also found 
in the first order formulation of 
these kinetic terms. 

\paragraph{Multiple spin-2 fields}
The natural candidates for the nonlinear, geometric completion for \eqref{ge-s2} are
\be
\mathcal L=\sum f R\wedge \dots \wedge R \wedge \o \wedge \dots \wedge \o \wedge E\wedge \dots \wedge E,
\ee
where $f$ are constants, $E$ could be different vielbeins and $R,\, \o$ could be different curvature two-forms and spin connections 
that are associated with different vielbeins. 

In general, curvature two-forms, spin-connections and vielbeins are the geometric forms in the general ghost-free Lagrangians \eqref{ge-df} 
in the formulation of differential forms.

\subsection{Coupled fields}\label{subs-coupled}
The constructions of ghost-free interaction among different kinds of fields are also straightforward, 
which can be done by using the building blocks in \eqref{o}. 
$f$ as a function should depend only on the scalar fields and the single-index terms. 
One should be careful about the single-index terms according to the discussion in subsection \ref{subs-sindex}. 

However, subtleties arise when we interpret the spin-2 fields as metric perturbations. 

Gravity is universally coupled to all matter. 
When we consider curved spacetime, the Minkowski metric and the Minkowski vielbein 
should be substituted with the curved metric and the dynamical vielbein. 
Derivatives are replaced with covariant derivatives, which introduce additional contributions from connection. 

From the perturbative point of view, after covariantization, 
ghost-free interactions that do not involve spin-2 fields 
are accompanied by higher order terms that contain spin-2 fields. 
To obtain second order equation of motion, counter-terms should be introduce to cancel 
the dangerous terms induced by connection \cite{counter-term}. 
For example, The most general scalar-tensor theories, the Horndeski theories \cite{Horndeski:1974wa}, 
that directly lead to second order equations of motion were rediscovered 
by covariantizing the Galileons.
 
We want to comment that the new scalar-tensor theories 
proposed in \cite{Gleyzes:2014dya} are still encompassed in this general framework, 
as the new terms reduce to the form of eq. \eqref{sc2} at the lowest order. 
As we discuss in the end of subsection \ref{subs-genL}, 
only the lowest order terms need total antisymmetrization. 
The related developments are due to the subtleties 
from covariantization or the nonlinear structure of the gravitational sector. 

\section{Conclusions and outlook}\label{sec-ccl}
To summarize, a general framework for Lagrangian field theories is developed . 
The basic assumptions are Lorentz-invariance and 
the absence of higher (than second) order terms in the equations of motion for the decomposed fields. 
The antisymmetric structure emerges as the solution to the local ghost-free equation \eqref{loc}. 
Using the antisymmetric Kronecker delta, 
the general Lagrangians \eqref{gf-Lag} are constructed and the building blocks are \eqref{o}. 
Then we reformulate this framework in the language of differential forms \eqref{ge-df}. 
The geometric nature of ghost-freeness is elucidated. 
A novel duality is discovered and a related concept of ``double form'' is proposed. 
All well-established Lagrangian field theories have natural formulation in this framework. 
Many new theories are constructed as well.  

To guarantee the Hamiltonian is bounded from below, 
it is not enough to show the equations of motion are of second orders. 
We are working on the Hamiltonian analysis of the new theories, 
for example some cases of the spin-1 fields. 

Recently, in \cite{DeFelice:2015hla}, massive gravity with standard kinetic term 
in triangular gauge was investigated. 
Lorentz-invariance is spontaneously broken by the internal frame and 
it was claimed that only two degrees of freedom are propagating in this theory.  
Along this line, the new kinetic terms \eqref{gr-kin} in the triangular gauge, 
which was first proposed in \cite{Li:2015izu}, have more promise to be ghost-free. 

The connection between ghost-instability and higher order equations of motion is not completely clear. 
In particular, recent developments in scalar-tensor theories \cite{Gleyzes:2014dya}  indicate there are some interesting subtleties. 

It will be interesting to find first-order formulation of the general framework. 
The second order Lagrangian are recovered by solving the equations of motion for the auxiliary fields. 
The case of the Galilean invariant theories has been done and they turn out to be Wess-Zumino terms \cite{Goon:2012dy} . 

Another important extension of the general framework is to include half-integer spin fields, 
which are the matter content in the standard model of particle physics. 
The reformulation of this framework in terms of differential forms should be a proper starting point. 
In addition, the concept of ``double form" might be related to spinor or tangent/cotangent bundles formulations. 


\acknowledgments
I want to thank F. Nitti for the suggestion of writing this long version of \cite{Li:2015vwa} to explain the details. 
My thanks also go to E. Babichev, C. Charmousis, X. Gao, E. Kiritsis, J. Mourad, V. Niarchos, K. Noui, R. Saito and D. Steer for useful comments or/and discussions.
I would like to thank C. de Rham, K. Hinterbichler, A. Matas, 
A. Solomon and A. Tolley for correspondence.

This work was supported in part by European Union's Seventh Framework Programme
under grant agreements (FP7-REGPOT-2012-2013-1) no 316165, the EU program ``Thales" MIS 375734
 and was also cofinanced by the European Union (European Social Fund, ESF) and Greek national funds through
the Operational Program ``Education and Lifelong Learning" of the National Strategic
Reference Framework (NSRF) under ``Funding of proposals that have received
a positive evaluation in the 3rd and 4th Call of ERC Grant Schemes".

\appendix
\section{An example of Ostrogradsky's instability}\label{app-Ost}
To have some concrete idea of the Ostrogradsky's instability discussed in subsection \ref{subs-Ost}, 
let us consider an example in which the Lagrangian of a scalar field $\phi$ contains a fourth order interaction. 
The Lagrangian reads
\be
\mathcal L=-\frac 1 2\pa_\m\phi\,\pa^\m\phi+\frac 1 2 \sigma \Lambda^{-2}\, \phi\,\Box^2  \phi\label{ostr-ex},
\ee
where $\sigma$ is the sign of the fourth order term, 
$\Box=\pa_\m\pa^\m$ is the d'Alembert operator 
and $\Lambda$ is some energy scale. 
This is a simplified version of the example in \cite{Creminelli:2005qk}  where the potential term in \eqref{ostr-ex} vanishes. 

The fourth order term can be transformed to lower order by introducing an auxiliary scalar field $\chi$
\be
\mathcal L=-\frac 1 2\pa_\m\phi\,\pa^\m\phi-\s\,\pa_\m\chi \,\pa^\m \phi-\frac 1 2 \s\Lambda^2\chi^2.
\ee
By integrating out the auxiliary field or solving the equation of motion for $\chi$, 
one recovers the original Lagrangian \eqref{ostr-ex}.

The kinetic terms are diagonalized by introducing another scalar field
\be
\Phi=\phi+\s \chi.
\ee 

The Lagrangian \eqref{ostr-ex} becomes
\be
\mathcal L=-\frac 1 2\pa_\m\Phi\,\pa^\m\Phi+\frac 1 2 \pa_\m \chi\,\pa^\m\chi-\frac 1 2 \s\Lambda^2\,\chi^2.
\ee
We can see the auxiliary field $\chi$ becomes dynamical and the two kinetic terms have opposite signs. 
One of them must have a wrong sign even if we change the original sign of the kinetic term for $\phi$. 

Let us follow the original convention in the Lagrangian \eqref{ostr-ex}. 
The scalar field $\chi$ corresponds to ghost-like degrees of freedom due to the wrong sign of $(\pa\chi)^2$. 
The Hamiltonian is not bounded from below as the kinetic energy of $\chi$ is negative. 

The sign of the fourth order term determines whether $\chi$ is a tachyon with negative mass squared. 
From the perspective of effective field theories, $\Lambda$ is the cutoff scale and 
the ghost-like degrees of freedom will be excited when the energy scale is larger than $\Lambda$.

\section{Linear theories of spin-2 fields}\label{app-lins2}
From subsection \ref{subs-Ost} to subsection \ref{subs-tenind}, 
we discuss the ghost-free constraints on Lorentz-invariant theories 
and the emergence of antisymmetric structure from Lorentz-invariance and ghost-freeness.
 
Now let us consider a concrete example: the linear theories of symmetric rank-2 tensor fields. 
Some discussions of this appendix overlap with 
the section 2.2 of de Rham's comprehensive review on massive gravity \cite{deRham:2014zqa} .

We assume that the linear theories of spin-2 fields are Lorentz-invariant and free of ghost-like degrees of freedom. 
In the discussions below, antisymmetric structure emerges as a consequence of these two assumptions. 
The unique ghost-free kinetic term turns out to be the linearized Einstein-Hilbert action 
and the unique ghost-free mass term requires the Fierz-Pauli tuning. 

Schematically, the general quadratic action of the linear theory reads
\be
\mathcal L=\pa h\,\pa h+ h\, h,
\ee
where the symmetric rank-2 tensor can be thought of as the metric perturbation $h_{\m\n}=g_{\m\n}-\eta_{\m\n}$.

For Lorentz-invariant theories, the indices are contracted by the Minkowski metric. The Lagrangian is
\ba
\mathcal L=
&&\,\pa_\m {{h_\r}^\r}\, \pa_\n h^{\m\n}
+\pa_\m {h^\m}_\r \,\pa_\n h^{\n\r}
+\pa_\m {h^\n}_\r \,\pa_\n h^{\m\r}
\nonumber\\
&&\,
+\pa_\m h_{\n\r} \,\pa^\m h^{\n\r}
+\pa_\m {h_\n}^\n\, \pa^\m {h_\r}^\r
+({h_\m}^\m)^2
+h_{\m\n} h^{\m\n},\nonumber\\
\ea
where the Lorentz-invariant terms should be multiplied by some coefficients. 
The second term and the third term are related by integrating by parts 
and we keep only the second term. 
The general Lorentz-invariant quadratic action reads
\ba
\mathcal L=
&&\,c_1\,\pa_\m {{h_\r}^\r} \pa^\n {h_\n}^{\m}
+c_2\,\pa_\m {h_\r}^\m \pa^\n {h_\n}^\r
\nonumber\\
&&\,
+\,c_3\,\pa_\m {h_{\r}}^\n \pa^\m {h_\n}^\r
+c_4\,\pa_\m {h_\r}^\r \pa^\m {h_\n}^\n
\nonumber\\
&&\,
+\,c_5\,{h_\m}^\m{h_\n}^\n
+c_6\,h_{\m\n} h^{\m\n},
\label{spin-2-c}
\ea
where the coefficients $c_1,\dots,c_6$ will be constrained by the requirement that the ghost-like degrees of freedom are absent. 

Now let us follow the approach in subsection \ref{subs-lon} and decompose the metric perturbation into transverse and longitudinal modes
\be
h_{\m\n}=h_{\m\n}^T+\pa_\m A_\n+\pa_\n A_\m+\pa_\m\pa_\n \Phi,
\ee
where 
\be
\pa^\m h_{\m\n}^T=\pa^\m A_\m=0.
\ee
In four dimensions, $(h_{\m\n}^T,\, A,\,\Phi)$ have $(6,\,3,\,1)$ independent variables. 
The total number of independent variables is 10, 
which is the same as that of a symmetric rank-2 tensor.

In terms of the decomposed fields, the quadratic Lagrangian \eqref{spin-2-c} reads
\ba
\mathcal L=&&\,
c_3\,\pa_\m {(h^T)_{\r}}^\n \pa^\m {(h^T)_{\n}}^\r
+c_4\,\pa_\m {(h^T)_\r}^\r \pa^\m {(h^T)_\n}^\n
\nonumber\\&&\,
+c_5\,{(h^T)_\m}^\m{(h^T)_\n}^\n
+c_6\,(h^T)_{\m\n} (h^T)^{\m\n}
\nonumber\\&&\,
+2c_6\,\pa_\m A_\n \pa^\m A^\n
+2c_5\, {(h^T)_\m}^\m(\Box\Phi)
\nonumber\\&&\,
+(c_2+2c_3)\, A_\r\, \Box^2 A^\r
-(c_1+2c_4)\,{(h^T)_\m}^\m (\Box^2 \Phi)
\nonumber\\&&\,
-(c_1+c_2+c_3+c_4)\,\Phi \,\Box^3 \Phi
+(c_5+c_6)\,\Phi\,\Box^2\Phi.
\nonumber\\
\ea
The terms in the last two line will lead to higher order equations of motion and 
ghost-like degrees of freedom will be propagating according to Ostrogradsky's theorem. 
To have second order equations of motion, 
we require the coefficients of these higher order terms vanish
\be
c_2+2c_3=c_1+2c_4=c_1+c_2+c_3+c_4=c_5+c_6=0.
\ee
Since there are four equations for $(c_1,\dots, c_6)$, only two coefficients are independent. 
One corresponds to the coefficient of the kinetic term and the other one is the coefficient of the mass term. 
We will parametrize the ghost-free theories by $c_3$ and $c_5$. 
The general ghost-free Lagrangian reads
\ba
\mathcal L=
&&\,c_3(2\,\pa_\m {{h_\r}^\r} \pa^\n {h_\n}^{\m}
-2\,\pa_\m {h_\r}^\m \pa^\n {h_\n}^\r
\nonumber\\
&&\,
+\,\pa_\m {h_{\r}}^\n \pa^\m {h_\n}^\r
-\,\pa_\m {h_\r}^\r \pa^\m {h_\n}^\n)
\nonumber\\
&&\,
+c_5(\,{h_\m}^\m{h_\n}^\n
-\,h_{\m\n} h^{\m\n}),
\ea
which can be written in a compact form using antisymmetrization
\be
\mathcal L=
c_3\, {h_\r}^{[\r} \pa_\m\pa^\m {h_\n}^{\n]}
+c_5\,{h_\m}^{[\m}{h_\n}^{\n]}.
\ee
Note that the indices in $[\dots]$ are antisymmetrized and here we use the unnormalized convention.
The first term is the linearized Einstein-Hilbert term and the second term is the Fierz-Pauli term. 
They are special cases of the general ghost-free Lagrangian 
for spin-2 fields discussed in subsection \ref{subs-exs2}.

In terms of the decomposed fields, the ghost-free Lagrangian is
\ba
\mathcal L=&&\,
c_3\left[\pa_\m {(h^T)_{\r}}^\n \pa^\m {(h^T)_{\n}}^\r
-\,\pa_\m {(h^T)_\r}^\r \pa^\m {(h^T)_\n}^\n\right]
\nonumber\\&&\,
+c_5\,\left[{(h^T)_\m}^\m{(h^T)_\n}^\n
-\,(h^T)_{\m\n} (h^T)^{\m\n}\right]
\nonumber\\&&\,
-2c_5\,\pa_\m A_\n\, \pa^\m A^\n
+2c_5\, {(h^T)_\m}^\m(\Box\Phi).\qquad
\ea

The linearized Einstein-Hilbert provides a kinetic term for the transverse tensor mode $h^T_{\m\n}$. 
In massless theories with $c_5=0$, the longitudinal modes do not appear in the Lagrangian, so the Lagrangian is invariant under arbitrary change of the longitudinal modes. In this case, the longitudinal modes correspond to the gauge freedom and the linear Lagrangian is invariant under linearized diffeomorphisms. 

When $c_5\neq 0$, the Fierz-Pauli mass term generates a Maxwell kinetic term for the vector modes $A_\m$. 
Following the analysis of spin-2 fields, one can show that the Maxwell kinetic term is precisely
the unique ghost-free kinetic term for spin-1 fields. 
Here it emerges from the ghost-free mass term for the spin-2 fields. 

A positive mass squared $m^2$ for $h^T_{\m\n}$ corresponds to $c_5>0$. 
Correspondingly, the kinetic term for the spin-1 field has a correct sign as well.

The crossing term $h^T \,(\Box \Phi)$ requires diagonalization. 
It can be shown that after diagonalization the longitudinal scalar mode $\Phi$ do not have a kinetic term.
The equation of motion for $\Phi$ is a constraint.
$\Phi$ is determined to be proportional to the trace of $h^T_{\m\n}$.
 
Below are the details of diagonalization. 
The kinetic terms can be diagonalized by introducing a spin-2 field
\be
H_{\m\n}=h^T_{\m\n}+C\,\Phi \,\eta_{\m\n},\quad C= {c_5}/({3 c_3})
\ee
and the Lagrangian of the decomposed fields is
\ba
\mathcal L=&&\,
c_3(\pa_\m {H_{\r}}^\n \pa^\m {H_{\n}}^\r
-\,\pa_\m {H_\r}^\r \pa^\m {H_\n}^\n)
\nonumber\\&&\,
+\,c_5(\,{H_\m}^\m{H_\n}^\n
-\,H_{\m\n} H^{\m\n})
\nonumber\\&&\,
+\,6C {H_\m}^\m\, \Phi+12 C^2 \Phi^2
\nonumber\\&&\,
-2c_5\,\pa_\m A_\n \pa^\m A^\n.
\ea

\section{Translating Galileons into the language of differential forms}\label{app-form}
In this appendix, we consider the theories of single scalar field and 
translate the general ghost-free terms from the language of tensor \eqref{gf-Lag} to that of differential form \eqref{ge-df}.

In the original formulation, the general ghost-free terms for single scalar field read
\be
\mathcal L_{\text{Galileon}}^{(k)}=d^Dx\,
f \,\delta_{\n_1\dots\n_{k}}^{\m_1\dots\m_{k}}\prod_{i=1}^k \pa_{\m_{i}}\pa^{\n_{i}}\Phi, 
\ee
where f are scalar functions of $\Phi$ and $\pa_\r\Phi\pa^\r\Phi$. 

In the first step, we write the antisymmetric Kronecker delta in terms of the product of Levi-Civita symbols using eq. \eqref{2Levi}
\ba
\mathcal L_{\text{Galileon}}^{(k)}=&&\,
d^Dx \,f \,\e_{\n_1\dots\n_{D}}\e^{\m_1\dots\m_{D}}
\nonumber\\&&\,\qquad\qquad
\left(\prod_{i=k+1}^D{\eta_{\m_i}}^{\n_i}\right)
\left(\prod_{i=1}^k \pa_{\m_{i}}\pa^{\n_{i}}\Phi\right) 
\nonumber\\
\ea
and replace the volume form $d^Dx$ and the $\m$-Levi-Civita symbols with the wedge products of $dx^{\m_i}$
\ba
\mathcal L_{\text{Galileon}}^{(k)}=
f \,\e_{\n_1\dots\n_{D}}\left(\prod_{i=1}^k \pa_{\m_{i}}\pa^{\n_{i}}\Phi\right)
\left(\prod_{i=k+1}^D{\eta_{\m_i}}^{\n_i}\right)
\nonumber\\
dx^{\m_1}\wedge\dots\wedge dx^{\m_D}. \quad 
\ea
In the second step, we use the Minkowski vielbein $\eta^\n={\eta_\m}^\n dx^\m$
\ba
\mathcal L_{\text{Galileon}}^{(k)}=&&\,
f \,\e_{\n_1\dots\n_{D}}\left(\prod_{i=1}^k \pa_{\m_{i}}\pa^{\n_{i}}\Phi\right)
\nonumber\\&&\,
dx^{\m_1}\wedge\dots\wedge dx^{\m_k}\wedge \eta^{\n_{k+1}}\wedge\dots\wedge \eta^{\n_{D}}. 
\ea

In the third step, we introduce a matter differential form
\be
\o^\n=\pa_\m\pa^\n\Phi\,dx^\m
\ee
and the Lagrangians become
\ba
\mathcal L_{\text{Galileon}}^{(k)}=f  \,\e_{\n_1\dots\n_{D}}\,
\o^{\n_1}\wedge\dots\wedge\o^{\n_k}\wedge\eta^{\n_{k+1}}\wedge\dots\wedge\eta^{\n_D}
\nonumber\\
\ea
or
\be
\mathcal L_{\text{Galileon}}=f\,\o\wedge\dots\wedge\o\wedge\eta\wedge\dots\wedge\eta.
\ee

\section{Examples of double forms}\label{app-douf}
To illustrate the concept of double form proposed in subsection \ref{subs-douform}, 
we reformulate the Lagrangians of two conventional theories in terms of double forms. 
In a double form, both the two Levi-Civita symbols are used to construct D-forms, so we have two of them.

Firstly, we consider the 4D Maxwell kinetic term 
\be
\mathcal L_{\text{Maxwell}}=F\wedge \ast F. 
\ee

As a double form, it reads
\be
\mathcal L_{\text{Maxwell}}\sim F\wedge \tilde F \wedge \eta\wedge \eta,
\ee
where $F$ is a $(2,\,0)$-form, $\tilde F$ is a $(0,\,2)$-form, $\eta$ is a $(1,1)$-form
\ba
F=&&\,\pa_{\m_1}A_{\m_2} (dx^{\m_1}\wedge dx^{\m_2})(\tilde I),
\\
\tilde F=&&\,\pa_{\n_1}A_{\n_2} (I)(d\tilde x^{\n_1}\wedge d\tilde x^{\n_2}),
\\
\eta=&&\,\eta_{\m\n}(dx^\m)(d\tilde x^\n).
\ea
The Maxwell kinetic  term in four dimensions is a $(4,\,4)$-form
\ba
\mathcal L_{\text{Maxwell}}\sim \pa_{\m_1}A_{\m_2}\pa_{\n_1}A_{\n_2} \eta_{\m_3\n_3}\eta_{\m_4\n_4}
(d^4 x)^{\m}(d^4 \tilde x)^{\n}
\nonumber\\
\ea
where
\ba
(d^4 x)^{\m}=(dx^{\m_1}\wedge dx^{\m_2}\wedge dx^{\m_3}\wedge dx^{\m_4}),\\
(d^4 \tilde x)^{\n}=(d\tilde x^{\n_1}\wedge d\tilde x^{\n_2}\wedge d\tilde x^{\n_3}\wedge d\tilde x^{\n_4}).
\ea

Secondly, let us consider the example of Einstein-Hilbert kinetic term in four dimensions. 
In the standard formulation, the Einstein-Hilbert Lagrangian reads
\be
\mathcal L_{\text{EH}}=\e_{\n_1\n_2\n_3\n_4}\,R^{\n_1\n_2}\wedge E^{\n_3} \wedge E^{\n_4},
\ee
where $\n_i$ are the indices of the internal Lorentz frame and 
the external (spacetime) differential form is constructed from  the $\m$-indices. 

Now we use the Levi-Civita symbol to construct one more D-form. 
The Einstein-Hilbert term becomes
\be
\mathcal L_{\text{EH}}=R_{\m_1\m_2,\,\n_1\n_2}\,E_{\m_3\n_3}\,E_{\m_4\n_4}\,(d^4x)^\m\, (d^4\tilde x)^\n,
\ee
where $\m_i$ are external indices and $\n_i$ are internal indices. 
$(d^4x)^\m$ is an external 4-form and $(d^4\tilde x)^\n$ is an internal 4-form. 
The $\m$-indices are raised and lowered by the external metric $g_{\m_1\m_2}$ 
and the $\n$-indices by the internal metric $\tilde g_{\n_1\n_2}$. 
We do not restrict the internal frame to be a Lorentz frame, 
so $\tilde g_{\n_1\n_2}$ is not necessarily a Minkowski metric. 

Now let us examine the building blocks of the Einstein-Hilbert term. 
A ``vielbein'' as a double form is a $(1,1)$ form
\ba
E_{\m\n}(dx^\m)(d\tilde x^\n)=&&\,\tilde g_{\n_1\n_2}(d\tilde x^{\n_1})(d\tilde x^{\n_2})
\nonumber\\
=&&\,ds^2
\nonumber\\
=&&\,g_{\m_1\m_2}(d x^{\m_1})(d x^{\m_2}),
\ea
which is related to the line element. 

Similarly, a ``curvature two-form" become a  $(2,2)$ form 
\ba
&&\,R_{\m_1\m_2,\,\n_1\n_2}(dx^{\m_1}\wedge dx^{\m_2})(d\tilde x^{\n_1}\wedge d\tilde x^{\n_2})
\nonumber\\
=&&\,
R_{\m_1\m_2,\,\m_3\m_4}(dx^{\m_1}\wedge dx^{\m_2})(d x^{\m_3}\wedge d x^{\m_4})
\nonumber\\
=&&\,
R_{\n_1\n_2,\,\n_3\n_4}(d\tilde x^{\n_1}\wedge d\tilde x^{\n_2})(d \tilde x^{\n_3}\wedge d\tilde  x^{\n_4}),
\ea
where $R$ are all different in the three lines and they are related by ${E_\m}^\n$ due to the changes of basis. 

Both ``vielbeins" and ``curvature two-forms" are self-dual in the formulation of double forms. 
In this way, we can see the Einstein-Hilbert term is a self-dual double form. 
More generally, the Lovelock terms are the wedge products of vielbeins and several curvature two-forms, 
so we can conclude that the Lovelock terms are self-dual as well. 

The Lovelock terms in the double form formulation are invariant 
under the changes of both internal and external frames. 
The standard Lagrangian in a scalar action is the component of the double form 
with one of the D-forms gauge-fixed to be the Minkowski vielbein. 
The reason behind this connection is not very clear.


\end{document}